\documentclass[prb,preprint]{revtex4-1}

\usepackage{amsmath}  
\usepackage{amsfonts} 
\usepackage{graphicx} 

\begin{document}


\title{Demonstrating two-particle interference with a one-dimensional delta potential well}

\author{Zhi Jiao Deng}
\email{dengzhijiao926@hotmail.com} 
\author{Xin Zhang}
\author{Yong Shen}
\author{Wei Tao Liu}
\author{Ping Xing Chen}
\affiliation{Institute for Quantum Science and Technology, College of Science, National University of Defense Technology, Changsha 410073, China}
\affiliation{Hunan Key Laboratory of Mechanism and Technology of Quantum Information, Changsha 410073, China}


\date{\today}

\begin{abstract}
In quantum mechanics, the exchange symmetry of wave functions for identical particles has observable effects, including the widely studied Hong-Ou-Mandel (HOM) effect. A theoretical description using second quantization is elegant but abstract. In contrast, this paper describes a simple model of two-particle interference using a one-dimensional delta potential well as a beam splitter. The conditions for the HOM effect are derived from the perspective of wave packet evolution. Furthermore, the interference processes of bosons, fermions and distinguishable particles are demonstrated and compared in detail. The method presented here is concrete, easy to visualize, and can help students to better understand the effects arising from the exchange symmetry of wave functions. The main results can be animated for classroom teaching or developed into an undergraduate seminar topic.
\end{abstract}

\maketitle 

\section{Introduction} 

Teaching about identical particles in quantum mechanics is both important and difficult \cite{singh1,	 singh2}. The wave functions of bosons and fermions have to be symmetric and antisymmetric, respectively, under the exchange of two particles \cite{Ehrenfest}. This requirement leads to many observable effects, such as the Pauli exclusion principle, exchange forces, degeneracy pressure, lasers, and Bose-Einstein condensation \cite{Griffiths, Huang}. Hong-Ou-Mandel (HOM) interference is another such effect \cite{zongshu}. It refers to the interference of two indistinguishable photons at a beam splitter. The HOM effect has been utilized in precision measurement \cite{measure1, measure2}, quantum state engineering \cite{engineering}, quantum computation \cite{computer}, and quantum communication \cite{communicate}. It has also been extended to electrons \cite{electron}, phonons \cite{phonon}, atoms \cite{atom}, and multi-particle interference \cite{multiparticle}.

As a two-particle interference phenomenon, the HOM effect clearly shows the differences between bosons, fermions and distinguishable particles. Therefore, it is an ideal teaching example, and can be found in undergraduate quantum physics courses with an experimental component\cite{course1, course2, course3}. The theoretical description is usually a relatively abstract second quantization method involving the creation and annihilation operators \cite{zongshu}. This approach is elegant but difficult for undergraduates who are new to quantum mechanics. From the perspective of time evolution, two-particle interference can be regarded as a process in which two wave packets move towards each other, meet and interfere at a beam splitter, and then separate. The goal of this manuscript is to present a simple and intuitive model of this process.

In this paper, the evolution of two-particle interference will be demonstrated via a one-dimensional delta potential well. The main purpose is to compare the behavior of bosons, fermions, and distinguishable particles, so as to help students to understand effects arising from the exchange symmetry of wave functions. The potential well plays the role of a beam splitter and the analysis only requires the one-dimensional Schr\"{o}dinger equation. It should be noted that although the model only applies to non-relativistic particles with mass, the physical pictures are consistent with photons. Moreover, wave packet dispersion is inherent in the state evolution, thus giving rise to new features compared with photons.

Similar work was recently published by Bermann and Rojo \cite{Bermann}. Before calculating scattering by a specific potential, those authors used optimal interference conditions to demonstrate the HOM effect through a wave packet mechanism. Using the one-dimensional delta potential barrier, they focused on the time evolution of the probabilities for two particles to appear on the same or opposite side of the barrier. Our study treats the same problem in a different manner and complements their work. First, optimal interference conditions are not set at the beginning, but are derived. Second, two kinds of probability densities have been defined to reveal more interference details. Both the joint probability distribution and the probability distribution for separation in real space vividly show how these two particles are related to each other. Wave packet dispersion is also included, and all possible deviations from the optimal conditions are checked individually to strengthen the understanding of two-particle interference.

The structure of this paper is as follows: Section II present a detailed analysis of delta potential well as a beam splitter and introduces our model of two-particle interference.  Analytical solutions of wave packet evolution are derived, and two probability distributions are defined and analyzed. Then, we derive optimal interference conditions by comparing probability distributions of bosons, fermions, and distinguishable particles. In Section III, scattering under optimal interference conditions and scattered final states under non-optimal conditions are compared. Section IV discusses how the model can be used in the classroom.

\section{MODEL AND ANALYTICAL DERIVATION}

An indispensable element in realizing the HOM effect is the beam splitter, which divides incident particles into two beams \cite{beamsplitter}. The one-dimensional delta potential well is a simple model that functions as a beam splitter in quantum scattering. Here, we demonstrate how this beam splitter works and how to change its splitting ratio.

\subsection{Delta potential well beam splitter}

Suppose the delta potential well $V(x) = - \alpha\delta(x)$ with depth parameter $\alpha > 0$ is located at the coordinate origin, and a particle with mass $m$ incident from the left is described by a normalized Gaussian wave packet at time $t=0$,
\begin{equation}
\Phi\left( {x,0} \right) = \left( \frac{2\mathrm{\Delta}}{\pi} \right)^{\frac{1}{4}}e^{- \mathrm{\Delta}{(x - s)}^{2}}e^{ik_{0}x},
\label{eq1}
\end{equation}
where $s < 0$ and $1/({2}\sqrt{\Delta})$ represent the initial average value and standard deviation of position, respectively, and $\hbar k_{0} > 0$ denotes the average incident momentum. The time-dependent evolution of this wave packet requires solving the Schr\"{o}dinger equation.

For a particle with definite energy $E$ incident from the left and $k = {\sqrt{2mE}/\hbar}> 0$, the stationary state solution reads \cite{Griffiths},
\begin{equation}
\psi_{k}\left( {x,t} \right) = \left\{
\begin{array}{ll}
   \left( e^{ikx} + \frac{i\beta(k)}{1 - i\beta(k)}e^{- ikx} \right)e^{- i\frac{\hbar k^{2}t}{2m}},~~\left( {x \leq 0} \right)\\
   \frac{1}{1 - i\beta(k)}e^{ikx}~e^{- i\frac{\hbar k^{2}t}{2m}},~~~~~~~~~~~~~~~~(x > 0)\\
\end{array}\right.\\
\label{eq2}
\end{equation}
where $\beta(k) = m\alpha/\left( \hbar^{2}k \right)$ is related to the well depth parameter $\alpha$ and wave vector $k$. Similarly, the solution for a particle incident from the right with the same definite energy but an opposite momentum $k = -{\sqrt{2mE}/\hbar}< 0$, takes the form,
\begin{equation}
\psi_{k}\left( {x,t} \right) = \left\{
\begin{array}{ll}
   \frac{1}{1 + i\beta(k)}e^{ikx}~e^{- i\frac{\hbar k^{2}t}{2m}},~~~~~~~~~~~~~~~~~\left( {x \leq 0} \right)\\
   \left( e^{ikx} - \frac{i\beta(k)}{1 + i\beta(k)}e^{- ikx} \right)e^{- i\frac{\hbar k^{2}t}{2m}},~~~~(x > 0)\\
\end{array}\right.\\
\label{eq3}
\end{equation}
with $\beta(k) = m\alpha/\left( \hbar^{2}k \right)< 0$ due to $k < 0$.
In applying Eqs.~(\ref{eq2}) and (\ref{eq3}), it is convenient to distinguish the incident directions by the sign of $k$ and use a single symbol $\psi_{k}\left( {x,t} \right)$ to denote the basis states.

These solutions in Eqs.~(\ref{eq2}), (\ref{eq3}) satisfy ${\int_{- \infty}^{\infty}{\psi_{k}^{*}(x,0)\psi_{k^{'}}\left( {x,0} \right)}}dx = 2\pi\delta\left( k - k^{'} \right)$, and thus form a complete and orthogonal basis. Using linear combinations of these solutions, it is straightforward to derive the wave packet solution. The key is to project the initial state onto the basis state and work out the superposition coefficient,
\begin{equation}
\phi(k) = \frac{1}{\sqrt{2\pi}}{\int_{- \infty}^{\infty}{\psi_{k}^{*}(x,0)\Phi\left( {x,0} \right)}}dx.
\label{eq4}
\end{equation}
If the initial state is localized on the left side of the well with $|s| \gg 1/\sqrt{\Delta}$, then
\begin{equation}
	\phi(k)= \begin{cases}\left(\frac{1}{2 \pi \Delta}\right)^{\frac{1}{4}}\left(e^{i\left(k_0-k\right) s} e^{-\frac{\left(k-k_0\right)^2}{4 \Delta}}-\frac{i \beta(k)}{1+i \beta(k)} e^{i\left(k+k_0\right) s} e^{-\frac{\left(k+k_0\right)^2}{4 \Delta}}\right), & (k>0) \\ \left(\frac{1}{2 \pi \Delta}\right)^{\frac{1}{4}}\left(\frac{1}{1-i \beta(k)} e^{i\left(k_0-k\right) s} e^{-\frac{\left(k-k_0\right)^2}{4 \Delta}}\right), & (k<0)\end{cases}
\label{eq5}
\end{equation}
and the wave packet solution can be obtained by substituting Eqs.~(\ref{eq2}), (\ref{eq3}), and (\ref{eq5}) into the expression:
\begin{equation}
\Phi\left( {x,t} \right) = \frac{1}{\sqrt{2\pi}}{\int_{- \infty}^{\infty}{\phi(k)}}\psi_{k}\left( {x,t} \right)~dk.
\label{eq6}
\end{equation}
In addition, if the average incident momentum is much larger than its uncertainty, i.e., $\ k_{0} \gg \sqrt{\mathrm{\Delta}}$, the solution only has contributions from the left incident basis states, so the wave packet solution simplifies to
\begin{equation}
\Phi\left( {x,t} \right) \approx \left\{
\begin{array}{ll}
   \frac{1}{\sqrt{2\pi}}{\int_{0}^{\infty}{\phi(k)}}\left( {e^{ikx} + \frac{i\beta(k)}{1 - i\beta(k)}e^{- ikx}} \right)~e^{- i\frac{\hbar k^{2}t}{2m}}dk,~~~(x \leq 0) \\
   \frac{1}{\sqrt{2\pi}}{\int_{0}^{\infty}{\phi(k)}}~\frac{1}{1 - i\beta(k)}e^{ikx}e^{- i\frac{\hbar k^{2}t}{2m}}~~dk,~~~~~~~~~~~~~~~~~(x > 0)\\
\end{array}\right.\\
\label{eq7}
\end{equation}
with the superposition coefficient approximated as,
\begin{equation}
\phi(k) \approx\left(\frac{1}{2 \pi \Delta}\right)^{\frac{1}{4}} e^{i\left(k_0-k\right) s} e^{-\frac{\left(k-k_0\right)^2}{4 \Delta}}.~~~~(k>0)
\label{eq8}
\end{equation}
This is just the Fourier transform of the initial state. The solution in Eq.~(\ref{eq7}) is a piecewise function. The incident and reflected components are on one side of the coordinate origin, while the transmitted components are on the other side.

The transmission coefficient and reflection coefficient for monochromatic plane waves are $T_\text{p}(k)=1/\left( 1 + \beta(k)^{2} \right)$ and $R_\text{p}(k)=\beta(k)^{2}/\left( 1 + \beta(k)^{2} \right)$ with $T_\text{p}(k) + R_\text{p}(k) = 1$.  After scattering, the incident wave packet is completely decomposed into reflected and transmitted wave packets. The total probability on the right side of the well is
\begin{equation}
P_\text{R}(t) = {\int_{0}^{\infty}\left| {\Phi(x,t)} \right|^{2}}dx,
\label{eq9}
\end{equation}
and the transmission coefficient for a wave packet can be defined through its long time limit:
\begin{equation}
T = {\lim\limits_{t\rightarrow\infty}{P_\text{R}(t)}} = {\lim\limits_{t\rightarrow\infty}{\int_{0}^{\infty}\left| {\Phi\left( {x,t} \right)} \right|^{2}}}dx.
\label{eq10}
\end{equation}
If $\phi(k)$ has negligible contribution from $k<0$, then
\begin{equation}
T \approx {\int_{0}^{\infty}{\left| {\phi(k)} \right|^{2}T_\text{p}(k)}}dk .
\label{eq11}
\end{equation}
The transmission coefficient for a wave packet is the average over all the monochromatic plane wave components.

To visualize the scattering process, either the exact solution Eq.~(\ref{eq6}) or approximate solution Eq.~(\ref{eq7}) can be plotted. The approximate solution works well when the superposition coefficient has a complete Gaussian distribution. It is convenient to take $1/\sqrt{\Delta}$ as the unit of length and introduce the following dimensionless quantities: position $X = \sqrt{\Delta} \cdot x$, wave vector $K = {k/\sqrt{\Delta}}$, time $\tau = \Delta\hbar t/(2m)$, initial average position $S = \sqrt{\Delta} \cdot s$, central wave vector $K_{0} = {k_{0}/\sqrt{\Delta}}$, and well depth parameter $\Lambda = \frac{m\alpha}{\hbar^{2}\sqrt{\Delta}}$. The function $\beta(k)$ can now be expressed as $\beta(k) = {\Lambda/K}$.

\begin{figure}[h!]
  \centering\includegraphics[width=9cm]{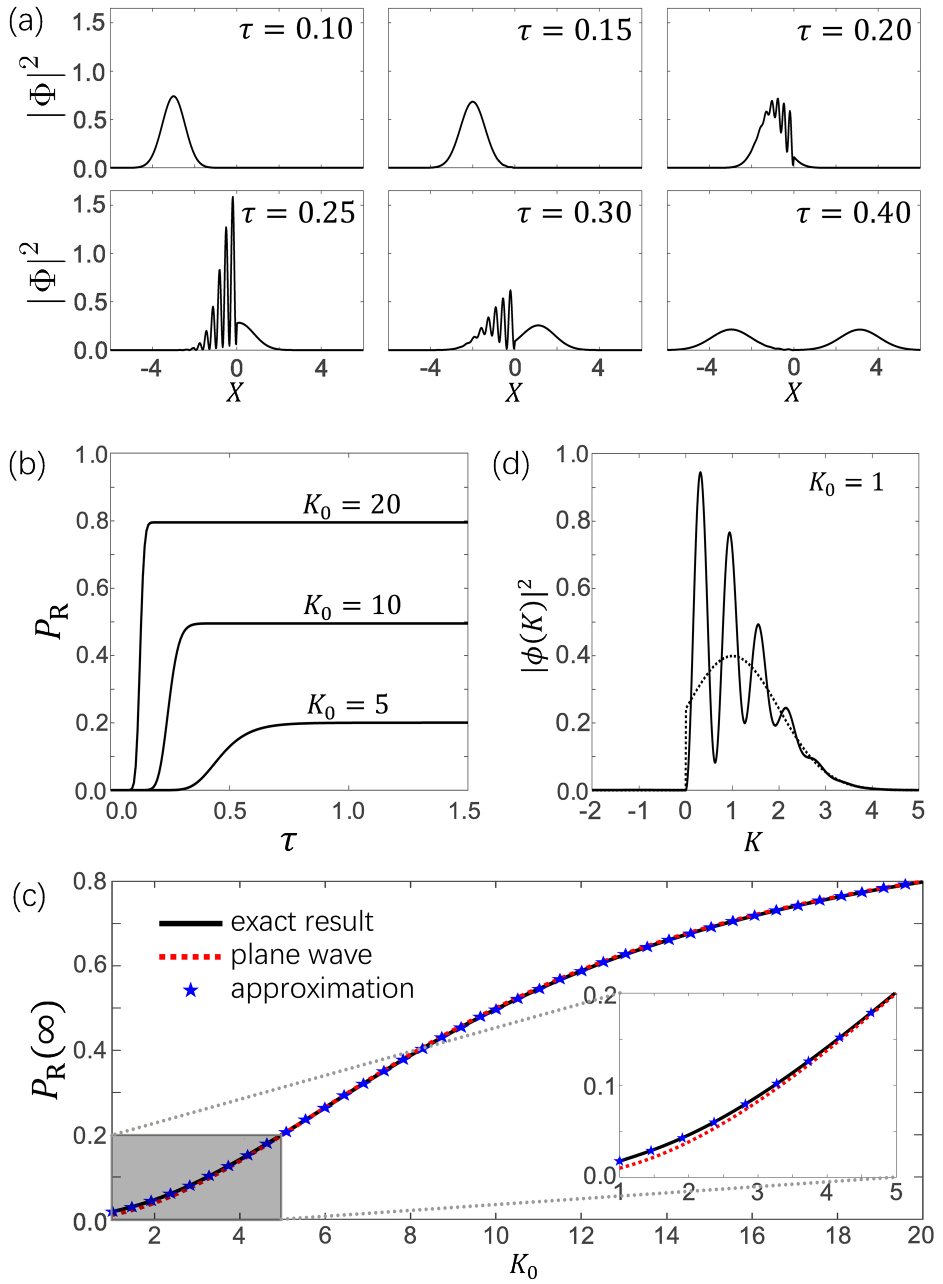}
  \caption{Delta function potential well as a beam splitter. (a) Scattering of Gaussian wave packet in the delta potential well with $K_{0} = 10$. (b) Time evolution of $P_\text{R}$, the probability to emerge on the right side of the well, for different values of $K_{0}$. (c) Wave packet transmission coefficient $T = P_\text{R}(\infty)$ as a function of central wave vector $K_{0}$. The black solid line and blue stars are obtained from numerical integration of the exact and approximate wave packet solutions, respectively. The red dotted line is the plane wave result: $T = 1/\left( 1 + \beta_{0}^{2} \right)$, where $\beta_{0} = {\Lambda/K_{0}}$. (d) The distribution of $\left| {\phi(K)} \right|^{2}$ for the exact (solid line) and approximate (dotted line) wave packet solutions with $K_{0} = 1$. Other parameters are $S = - 5$, $\Lambda = 10$, $\Delta = 1$.}
	\label{Figure1}
\end{figure}

As shown in Fig.~\ref{Figure1}(a), when the front of the incident wave packet reaches the location of the potential well at $X = 0$, the reflected wave component is immediately generated, resulting in interference fringes formed by the superposition of the incident and reflected wave packets in the region $X < 0$. After some time, the incident wave packet is fully transformed into two wave packets moving in opposite directions. The time evolution of $P_\text{R}$ in Fig.~\ref{Figure1}(b) shows that the probability saturates to a constant value, which is the transmission coefficient. The higher the incident energy, the greater the transmission coefficient. Under the assumption that $\ k_{0} \gg \sqrt{\mathrm{\Delta}}$, $\left| {\phi(k)} \right|^{2}$ is a complete Gaussian function with mean value $k_{0}$, so the value of $\beta(k)$ at central wave vector $\beta_{0} = \frac{m\alpha}{\hbar^{2}k_{0}} = {\Lambda/K_{0}}$ determines the wave packet transmission coefficient $T = 1/\left( 1 + \beta_{0}^{2} \right)$. The 50:50 beam splitter is realized at $\beta_{0} = 1$. To change the splitting ratio is to simply change the ratio of $\Lambda/K_{0}$.

If $k_{0}$ is comparable or smaller than $\sqrt{\mathrm{\Delta}}$, then $\left| {\phi(k)} \right|^{2}$ is a truncated Gaussian function in the approximate result of Eq.~(\ref{eq8}), while it exhibits oscillations in the exact result of Eq.~(\ref{eq5}), as shown in Fig.~\ref{Figure1}(d). These two distribution curves give rise to different state evolutions, but make negligible difference in the averaged transmission coefficients. They both lead to transmission coefficients slightly higher than that of a plane wave at the central wave vector, as shown in Fig.~\ref{Figure1}(c). On the whole, the formula $T = 1/\left( 1 + \beta_{0}^{2} \right)$ is a good estimate for the wave packet transmission coefficient.

Codes for all the subplots are accessible via the online supplementary material \cite{SM}.

\subsection{Two-particle interference}

Here, we introduce two tools for analyzing scattering: the joint probability distribution and the probability distribution for separation. the probability that both particles emerge from the same side of the well has also been derived and analyzed, from which the optimal interference conditions can be obtained.

\begin{figure}[h!]
  \centering\includegraphics[width=9cm]{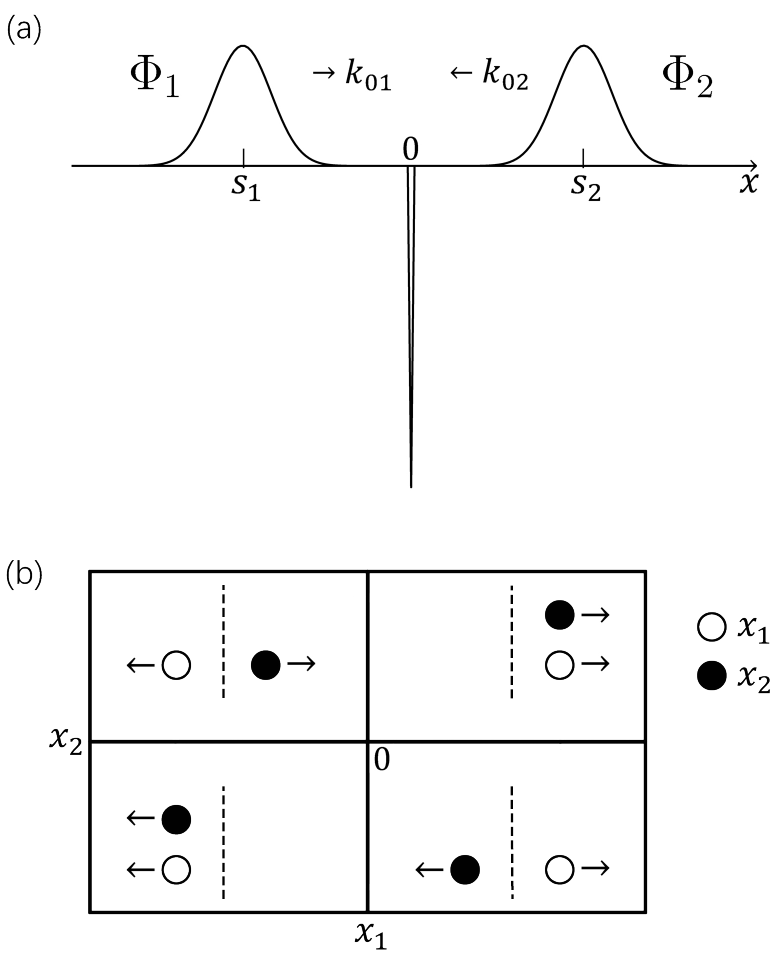}
  \caption{(a) Schematic diagram of two-particle interference. In a one-dimensional delta potential well, two incident particles with the same mass and described by Gaussian wave packets are incident from the left and right on a potential well at the origin. (b) Four outgoing channels after scattering. Particles 1 and 2, with their positions labeled by $x_{1}$, $x_{2}$, are represented by hollow and filled circles, respectively. In the $x_{1}-x_{2}$ two-dimensional position space, each quadrant represents a possible scattering channel corresponding to particles 1 and 2 located either to the left or to the right of the potential well, which is represented by a dashed line.}
	\label{Figure2}
\end{figure}

The two-particle interference model is shown in Fig.~\ref{Figure2}(a), where two particles with mass $m$ are incident from the left and right sides of the delta potential well, respectively. The states of the incident particles at $t=0$ can be written as,
\begin{equation}
  \begin{split}
     \Phi_{1}\left( {x,0} \right) &= \left( \frac{2\mathrm{\Delta}}{\pi} \right)^{1/4}e^{- \mathrm{\Delta}{({x - s_{1}})}^{2}}e^{ik_{01}x}\\
     \Phi_{2}\left( {x,0} \right) &= \left( \frac{2\mathrm{\Delta}}{\pi} \right)^{1/4}e^{- \mathrm{\Delta}{({x - s_{2}})}^{2}}e^{ik_{02}x},
  \end{split}
  \label{eq12}
\end{equation}
with initial average position $s_{1} < 0$, $s_{2} > 0$, and average incident momentum $\hbar k_{01} > 0$, $\hbar k_{02} < 0$. If there is no inter-particle interaction like attraction or repulsion, the two particles will be scattered independently in the potential well. They meet at the well, interfere due to wave packet overlap, convert into reflected and transmitted components, and then leave the scattering region. If the initial wave packets are well-localized on one side of the barrier ($\left| s_{j} \right| \gg 1/\sqrt{\Delta}$), and the average incident momentum is much larger than its uncertainty ($\left| k_{0j} \right| \gg \sqrt{\mathrm{\Delta}}$), then the wave packets evolve as follows,
\begin{equation}
 \begin{split}
   \Phi_{1}\left( {x,t} \right) &= \left\{
   \begin{array}{ll}
   \frac{1}{\sqrt{2\pi}}{\int_{0}^{+ \infty}{\phi_{1}(k)\left( {e^{ikx} + \frac{i\beta(k)}{1 - i\beta(k)}e^{- ikx}} \right)e^{- i\frac{\hbar k^{2}t}{2m}}}}dk,(x < 0)\\
   \frac{1}{\sqrt{2\pi}}{\int_{0}^{+ \infty}{\phi_{1}(k)\frac{1}{1 - i\beta(k)}e^{ikx}e^{- i\frac{\hbar k^{2}t}{2m}}}}dk,~~~~~~~~~~~~~~~(x \geq 0)\\
   \end{array}\right.\\
   \Phi_{2}\left( {x,t} \right) &= \left\{
   \begin{array}{ll}
   \frac{1}{\sqrt{2\pi}}{\int_{- \infty}^{0}{\phi_{2}(k)\frac{1}{1 + i\beta(k)}e^{ikx}e^{- i\frac{\hbar k^{2}t}{2m}}}}dk,~~~~~~~~~~~~~~~(x < 0)\\
   \frac{1}{\sqrt{2\pi}}{\int_{- \infty}^{0}{\phi_{2}(k)\left( {e^{ikx} - \frac{i\beta(k)}{1 + i\beta(k)}e^{- ikx}} \right)e^{- i\frac{\hbar k^{2}t}{2m}}}}dk, (x \geq 0)\\
   \end{array}\right.,\\
 \end{split}
 \label{eq13}
\end{equation}
with superposition coefficients $\phi_{j}(k) = \left( \frac{1}{2\mathrm{\Delta}\pi} \right)^{\frac{1}{4}}e^{i{({k_{0j} - k})}s_{j}}e^{- \frac{{(k - k}_{0j})^{2}}{4\mathrm{\Delta}}}, (j = 1,2)$. The wave packet transmission coefficients are $T_{j} = 1/\left( 1 + \beta_{0j}^{2} \right)$ with $\beta_{0j} = \frac{m\alpha}{\hbar^{2}k_{0j}}$, $(j = 1,2)$.

Since there is no inter-particle interaction, the two-particle wave function for distinguishable particles can be written in the form of a direct product state. However, if the particles are identical bosons or fermions, the wave function has to be symmetrized and anti-symmetrized, respectively.
\begin{equation}
	\begin{split}
		\Psi_{+}\left( x_{1},x_{2},t \right) &= \frac{1}{\sqrt{2}}\left( \Phi_{1}\left( {x_{1},t} \right)\Phi_{2}\left( {x_{2},t} \right) + \Phi_{1}\left( {x_{2},t} \right)\Phi_{2}\left( {x_{1},t} \right) \right) \\
		\Psi_{-}\left( x_{1},x_{2},t \right) &= \frac{1}{\sqrt{2}}\left( \Phi_{1}\left( {x_{1},t} \right)\Phi_{2}\left( {x_{2},t} \right) - \Phi_{1}\left( {x_{2},t} \right)\Phi_{2}\left( {x_{1},t} \right) \right) \\
		\Psi_\text{D}\left( {x_{1},x_{2},t} \right) &= \Phi_{1}\left( x_{1},t \right)\Phi_{2}\left( x_{2},t \right). \\
	\end{split}
\label{eq14}
\end{equation}
$\Psi_{+}$, $\Psi_{-}$, $\Psi_\text{D}$ represent the two-particle wave functions for bosons, fermions and distinguishable particles, respectively, and the positions of particles 1 and 2 are labeled by $x_{1}$, $x_{2}$.  All scenarios can be achieved via cold atoms \cite{cold}: two sodium-23 or rubidium-87 atoms are identical bosons, two lithium-6 or potassium-40 atoms are identical fermions, and two isotopes with nearly equal mass such as rubidium-87 and rubidium-85 are distinguishable particles.

Each scattered particle has some probability of being on the left or right side of the well, so there are four outgoing channels corresponding to the four quadrants in Fig.~\ref{Figure2}(b). The quadrants have a clear physical meaning: the first and third quadrants indicate that both particles emerge on the same side of the potential well, while the second and fourth quadrants indicate that they emerge on opposite sides. The symmetry of the wave function under exchange will lead to differences in the joint probability distribution $\left| \Psi_{i} \right|^{2}$ in position space.

Because of the indistinguishable nature of identical particles, it no longer makes sense to number the particles by 1 and 2.  Thus, we can introduce the variables $r = x_{1} - x_{2}$ and $R = \left( x_{1} + x_{2} \right)/2$, where $r$ and $R$ are the relative and center of mass coordinates, respectively. The probability distribution for separation of particles in the state of $\Psi_{i}$ ($i = +,-,\text{D}$) is
\begin{equation}
P_\text{sep}\left( |r|,t \right)\  = \ {\int_{- \infty}^{+ \infty}{\mid \Psi_{i}\left( R,|r|,t \right) \mid}^{2}}\ dR + {\int_{- \infty}^{+ \infty}{\mid \Psi_{i}\left( R, - |r|,t \right) \mid}^{2}}dR.
\label{eq15}
\end{equation}
This definition involves the summation of relative coordinates with opposite signs and the integration over the center of mass coordinates. It gives the statistical description of the inter-particle distance, and is closely linked to the question of whether the two particles stay together after the scattering.

Due to the amplitude interference of two-particle states, bosons tend to come out of the same port, and fermions tend to come out of different ports, while distinguishable particles do not have such a tendency \cite{zongshu}. After scattering, the single-particle wave functions can be fully decomposed into the superposition of the reflected component $\Phi_{j}^\text{R}$ and the transmitted component $\Phi_{j}^\text{T}$ as follows ($j = 1, 2$),
\begin{equation}
	\begin{split}
		\Phi_{1}\left( {x,t} \right) &= i\Phi_{1}^\text{R}(x,t) + \Phi_{1}^\text{T}(x,t) \\
		\Phi_{2}\left( {x,t} \right) &= i\Phi_{2}^\text{R}(x,t) + \Phi_{2}^\text{T}(x,t), \\
	\end{split}
\label{eq16}
\end{equation}
with
\begin{equation}
	\begin{split}
    \Phi_{1}^\text{R}\left( {x,t} \right) &= \frac{1}{\sqrt{2\pi}}{\int_{0}^{+ \infty}{\phi_{1}(k)\frac{\beta(k)}{1 - i\beta(k)}e^{- ikx}e^{- i\frac{\hbar k^{2}t}{2m}}dk, (x < 0)}} \\
	\Phi_{1}^\text{T}\left( {x,t} \right) &= \frac{1}{\sqrt{2\pi}}{\int_{0}^{+ \infty}{\phi_{1}(k)\frac{1}{1 - i\beta(k)}e^{ikx}e^{- i\frac{\hbar k^{2}t}{2m}}}}dk, (x > 0) \\
	\Phi_{2}^\text{R}\left( {x,t} \right) &= \frac{1}{\sqrt{2\pi}}{\int_{- \infty}^{0}{\phi_{2}(k)\frac{- \beta(k)}{1 + i\beta(k)}e^{- ikx}e^{- i\frac{\hbar k^{2}t}{2m}}}}dk, (x > 0)\\
	\Phi_{2}^\text{T}\left( {x,t} \right) &= \frac{1}{\sqrt{2\pi}}{\int_{- \infty}^{0}{\phi_{2}(k)\frac{1}{1 + i\beta(k)}e^{ikx}e^{- i\frac{ \hbar k^{2}t}{2m}}dk, (x < 0)}}. \\
	\end{split}
\label{eq17}
\end{equation}
These four wave packets will move farther away from the potential well over time, and their shape gets lower and wider due to dispersion. For the case that both particles come out from the right side, the particle incident from the left must be transmitted and the particle incident from the right must be reflected. Thus the probability amplitude of two identical particles appearing on the right side of the potential well is,
\begin{equation}
\Psi_\text{right} = \frac{i}{\sqrt{2}}\left\lbrack \Phi_{1}^{T}\left( {x_{1},t} \right)\Phi_{2}^{R}\left( {x_{2},t} \right) \pm \Phi_{1}^{T}\left( {x_{2},t} \right)\Phi_{2}^{R}\left( {x_{1},t} \right) \right\rbrack,
\label{eq18}
\end{equation}
where the $+$ and $-$ signs correspond to bosons and fermions, respectively. Combined with Eq.~(\ref{eq17}), the integration of the probability density $
\left| \Psi_\text{right} \right|^{2}$ in the entire coordinate space results in the probability that both particles appear on the right side, i.e.,
\begin{equation}
\begin{split}
	P_\text{right} &= {\iint_{- \infty}^{\infty}\left| \Psi_\text{right} \right|^{2}}dx_{1}dx_{2}\\
	&\approx \frac{{\beta_{02}}^{2}}{\left( {1 + {\beta_{01}}^{2}} \right)\left( {1 + {\beta_{02}}^{2}} \right)} \mp \frac{\beta_{01}\beta_{02}}{\left( {1 + {\beta_{01}}^{2}} \right)\left( {1 + {\beta_{02}}^{2}} \right)}e^{- \mathrm{\Delta}(s_{1} + s_{2})^{2}}e^{- \frac{(k_{01} + k_{02})^{2}}{4\mathrm{\Delta}}},\\
\end{split}
\label{eq19}
\end{equation}
where the second line adopts the assumption that the wave packet has a narrow range of frequency: i.e., $\left| k_{0j} \right| \gg \sqrt{\mathrm{\Delta}}$. Apparently, the wave packets evolve with time, however, the final scattering probability does not.

Adding the probability of both particles coming out from the left side, the total probability of detecting both particles on the same side is,
\begin{equation}
	\begin{split}
		P_{\pm} &\approx \frac{{\beta_{01}}^{2} + {\beta_{02}}^{2}}{\left( {1 + {\beta_{01}}^{2}} \right)\left( {1 + {\beta_{02}}^{2}} \right)} \mp \frac{2\beta_{01}\beta_{02}}{\left( {1 + {\beta_{01}}^{2}} \right)\left( {1 + {\beta_{02}}^{2}} \right)}e^{- \mathrm{\Delta}(s_{1} + s_{2})^{2}}e^{- \frac{(k_{01} + k_{02})^{2}}{4\mathrm{\Delta}}} \\
		&= P_\text{D} \pm 2\sqrt{T_{1}R_{1}T_{2}R_{2}}e^{- \mathrm{\Delta}(s_{1} + s_{2})^{2}}e^{- \frac{(k_{01} + k_{02})^{2}}{4\mathrm{\Delta}}}, \\
	\end{split}
\label{eq20}
\end{equation}
where $T_{j}$, $R_{j} = 1 - T_{j}$ are the transmission and reflection coefficients for the left ($j=1$) and right ($j=2$) incident particles, respectively. The first term $P_\text{D} = T_{1}R_{2} + R_{1}T_{2}$ is the result for distinguishable particles, and the second term arises from the exchange symmetry of the identical particles. Due to probability conservation, the probability of two particles emerging from opposite sides equals $(1 - P_{+ , - ,\text{D}})$, which is the coincidence probability measured in HOM experiments \cite{zongshu}. To make a significant difference, the second term should have a contribution as large as possible. This requires $T_{1} \cong T_{2} \cong 0.5$, $s_{1} + s_{2} \cong 0$, $k_{01} + k_{02} \cong 0$, which means that the two incident particles have the same energy and arrive at a 50:50 beam splitter simultaneously.

We can also analyze spin or polarization degrees of freedom. For simplicity, suppose that the dimension of spin space is 2, and the spin states of the incident particles from the left and right sides are $\left| \left. u \right\rangle \right.$ and $\left. c \middle| \left. u \right\rangle + d \middle| \left. v \right\rangle \right.$, respectively, with $|c|^{2} + |d|^{2} = 1$ and $\left| \left. u \right\rangle \right.$, $\left| \left. v \right\rangle \right.$ being orthonormal. Accordingly, the probability amplitude of detecting two particles on the right side becomes,
\begin{equation}
\begin{split}
\Psi_\text{right} = \frac{i}{\sqrt{2}}\left\{ \Phi_{1}^\text{T}\left( {x_{1},t} \right)\Phi_{2}^\text{R}\left( {x_{2},t} \right) \middle| \left. {u(1)} \right\rangle \otimes \left\lbrack c\left| {\left. {u(2)} \right\rangle + d} \right|\left. {v(2)} \right\rangle \right\rbrack \right.\\
\left. \pm \Phi_{1}^\text{T}\left( {x_{2},t} \right)\Phi_{2}^\text{R}\left( {x_{1},t} \right) \middle| \left. {u(2)} \right\rangle \otimes \left\lbrack {c\left| {\left. {u(1)} \right\rangle + d} \right|\left. {v(1)} \right\rangle} \right\rbrack \right\}.
\end{split}
\label{eq21}
\end{equation}
Analyzing the measurement probability in position space requires summing over the spin degrees of freedom, so there is
\begin{equation}
	P_\text{right} = {\iint_{- \infty}^{\infty}{\sum\limits_{l,n = u,v}\left\langle {\left. {\Psi_\text{right}\left| {l(1)n(2)} \right.} \right\rangle\left\langle {\left. {l(1)n(2)} \right|\Psi_\text{right}} \right.} \right\rangle}}dx_{1} dx_{2}.
\label{eq22}
\end{equation}
The final impact on the total probability is to multiply the overlap coefficient $|c|^{2}$ in the second term, i.e.,
\begin{equation}
	P_{\pm} \approx P_\text{D} \pm 2|c|^{2}\sqrt{T_{1}R_{1}T_{2}R_{2}}e^{- \mathrm{\Delta}(s_{1} + s_{2})^{2}}e^{- \frac{(k_{01} + k_{02})^{2}}{4\mathrm{\Delta}}}.
\label{eq23}
\end{equation}
For $c = 1$, the spin states are the same, and the interference effect is at its strongest, while for $c = 0$, the spin states are orthogonal, and there is no interference effect. In summary, to achieve the best comparison effect, the particles should also have the same spin state. These are the conditions under which the HOM effect occurs \cite{zongshu} and are here obtained entirely from the wave packet solution. For a simple check, with the optimal interference conditions, i.e., $T_{1} = T_{2} = 0.5$, $s_{1} + s_{2} = 0$, $k_{01} + k_{02} = 0$, $c = 1$, then $P_{+} = 1$, $P_{-} = 0$, and $P_\text{D} = 0.5$. Deviation from these conditions will weaken, or remove the observation effect. As shown in  Eq.~(\ref{eq23}), the influence from the differences in spin state is quadratic, while that from differences in incident energy and initial distance is exponential decay.

\section{COMPARISON AND ANALYSIS}

Here, we compare the interference for bosons, fermions and distinguishable particles using Eqs.~(\ref{eq13}) and (\ref{eq14}). The dimensionless quantities have the same values as in Fig.~\ref{Figure1}.

\begin{figure}[h!]
	\centering\includegraphics[width=15cm]{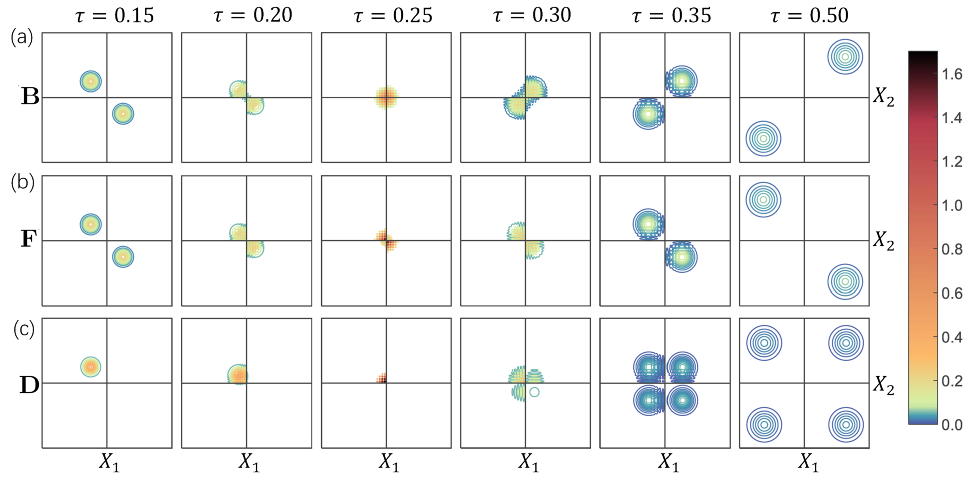}
	\caption{Comparison of the joint probability distributions of bosons, fermions and distinguishable particles in position space during the interference process. Each row labeled by (a), (b), (c) represents the joint probability densities $\left| \Psi_{+} \right|^{2}$, $\left| \Psi_{-} \right|^{2}$, $\left| \Psi_\text{D} \right|^{2}$ of bosons (\textbf{B}), fermions (\textbf{F}) and distinguishable particles (\textbf{D}) at different times, respectively. The parameters are set to achieve perfect HOM interference with $\Delta = 1$, $K_{01} = - K_{02} = 10$, $S_{1} = - S_{2} = - 5$, $\Lambda = 10$, and $c = 1$.}
	\label{Figure3}
\end{figure}

\begin{figure}[h!]
	\centering\includegraphics[width=9cm]{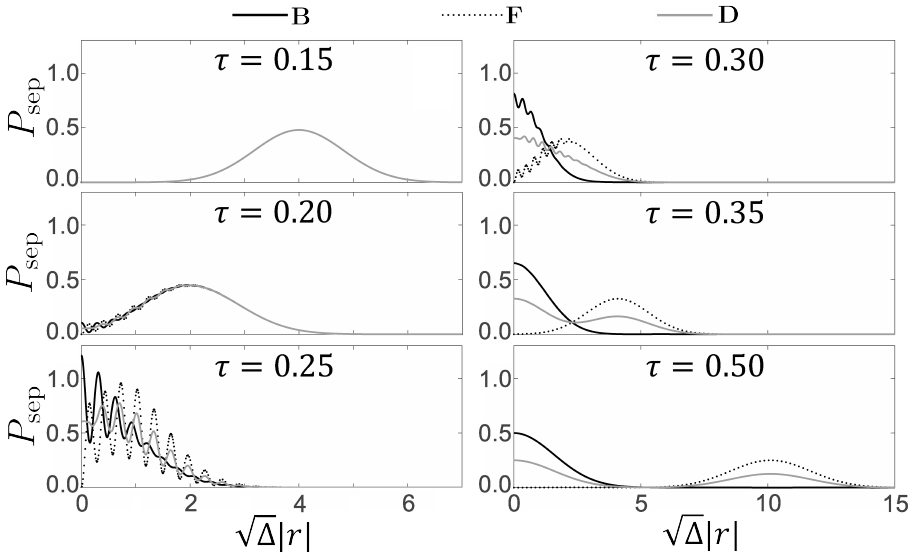}
	\caption{Comparison of the probability distributions for separation of bosons, fermions, and distinguishable particles at the times depicted in Fig.~\ref{Figure3}.}
	\label{Figure4}
\end{figure}

First, we compare scattering under HOM interference conditions. In Fig.~\ref{Figure3}, the two incident particles have the same energy, spin state, and initial distance from the well, and the incident energy is adjusted to achieve a 50:50 splitting ratio. Before scattering, the joint probability distribution of distinguishable particles in the position space has only one localized distribution, while that of identical particles has two because of the exchange of coordinates. At the initial stage, two particles with opposite coordinates move toward the origin. When their wave packets reach the well, interference between the incident and reflected components appears. Most crucially, the interference between the two localized distributions for identical particles causes the HOM effect. As shown in the last column of Fig.~\ref{Figure3}(a)(b)(c), the scattering of distinguishable particles has four outgoing channels with equal probability, while bosons and fermions only retain the channels for the same side and opposite sides, respectively.

Figure \ref{Figure4} shows the probability distribution for separation of particles in Eq.~(\ref{eq15}) corresponding to Fig.~\ref{Figure3}. When the wave packets have not yet reached the well, the distributions are exactly the same. When they reach the well, oscillatory fringes appear. The higher the overlap of the two incident wave packets, the more obvious the oscillation. Especially at $|r| = 0$, the probability density of fermions is 0, illustrating the Pauli exclusion principle \cite{Griffiths}, while the corresponding value for bosons is the largest, reflecting that bosons tend to stay together. After scattering back to the initial position, bosons and fermions have a single peak at $|r| = 0$ and $\sqrt{\Delta} \cdot |r| = 2\left| S_{1} \right|$, respectively. Meanwhile, distinguishable particles have peaks at both locations, forming a bimodal structure. The area of the peaks indicates that the probabilities of being on the same or opposite sides of the well are both $0.5$.

\begin{figure}[h!]
	\centering\includegraphics[width=9cm]{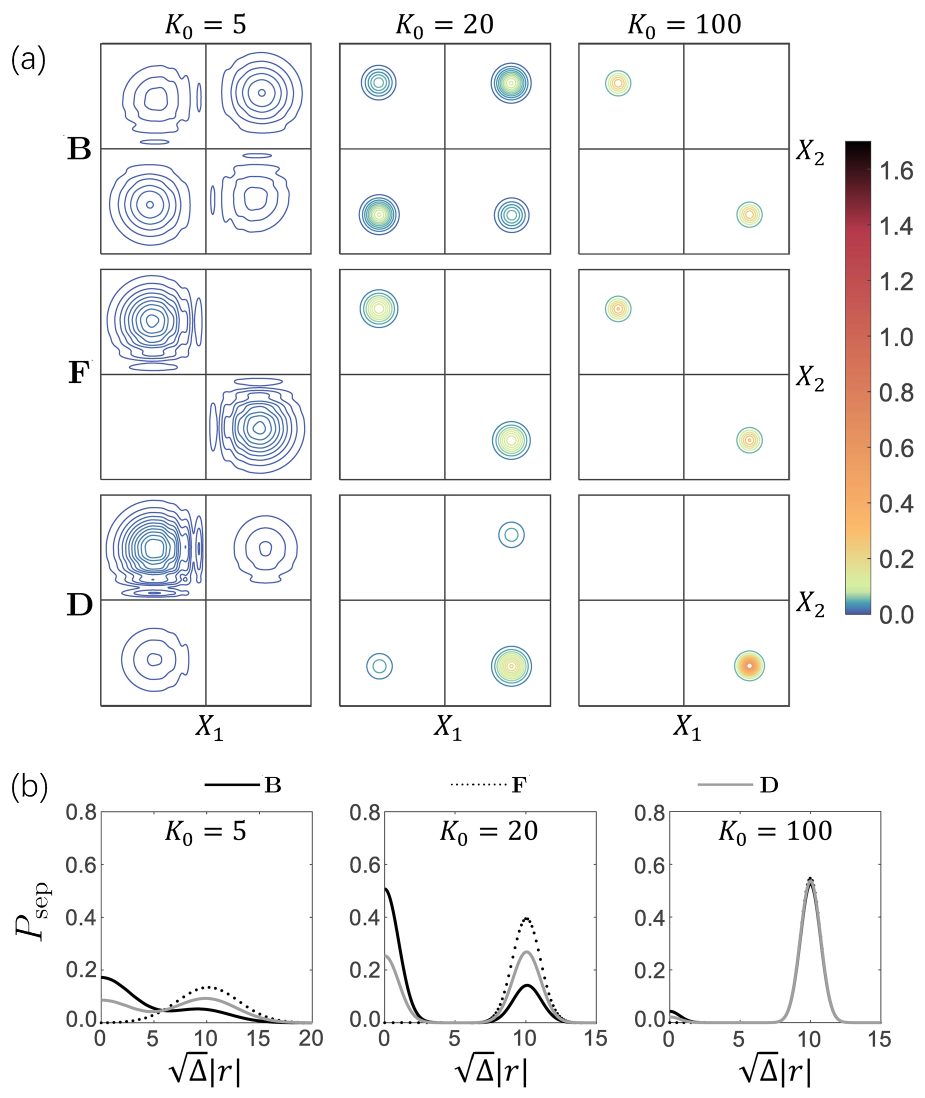}
	\caption{The scattered final states of bosons, fermions and distinguishable particles at different splitting ratios when the particles are scattered back to the initial position. (a) Each row from top to bottom represents the joint probability distributions of bosons (B), fermions (F), and distinguishable particles (D) in position space at different incident energies. The incident energy increases from left to right with $K_{01} = - K_{02} = K_{0}$, resulting in a sequential increase in the transmission coefficients $T_{1} = T_{2} = T = 0.2,0.8,0.99$. Other parameters are the same as in Fig.~\ref{Figure3}.  (b) Probability distributions for particle separation corresponding to (a).}
	\label{Figure5}
\end{figure}

The influence of the splitting ratio on scattering can be analyzed, too. The splitting ratio gives the ratio of transmission and reflection coefficients. It is related to the energies of the incident particles. In Fig.~\ref{Figure3}, the energies of the two incident particles are adjusted but kept equal so that $T_{1} = T_{2} = T$. According to Eq.~(\ref{eq23}), this will result in $P_{+} = 4T(1 - T)$, $P_{-} = 0$, and $P_\text{D} = P_{+}/2$.  The scattering of fermions are independent of the splitting ratio, and two identical fermions always emerge from opposite sides. The probability of bosons emerging together is enhanced by a factor of $2$ compared with that of the distinguishable particles, and when $T = 0.5$, $P_{+} = 1$. At $T = 0$ and $T = 1$, the probability of two particles being detected on the same side in all three cases is zero. In these two extreme cases, the particles are either completely reflected or transmitted, so they can only separate after scattering.

Figure \ref{Figure5} compares the scattered final states of bosons, fermions and distinguishable particles at different splitting ratios. Indeed, with $T = 0.99$ approaching 1, there only exist prominent distributions of two particles with opposite coordinates in position space, and the probability distributions for separation are almost the same in all three cases. The two particles simply meet and then separate. $T = 0.2,~0.8$ are selected to differ symmetrically from $T = 0.5$, with $P_{+}(0.2) = P_{+}(0.8)$, $P_\text{D}(0.2) = P_\text{D}(0.8)$. The splitting ratios $20:80$ and $80:20$ are reciprocal to each other, and the joint distributions for identical particles are quite similar in position space due to the exchange of coordinates. In particular, the coordinates for fermions are always opposite and the probability distribution for separation maintains a unimodal structure. By contrast, the joint distributions for distinguishable particles differ because the probability of both particles being reflected is greater at lower incident energy. In all cases, when the incident energy is smaller, the scattering process takes longer, and so wave packet dispersion is more obvious. The first two subplots in Fig.~\ref{Figure5}(b) have similar shapes, but for lower incident energy, the peaks become shorter and wider, and even merge together.

\begin{figure}[h!]
	\centering\includegraphics[width=9cm]{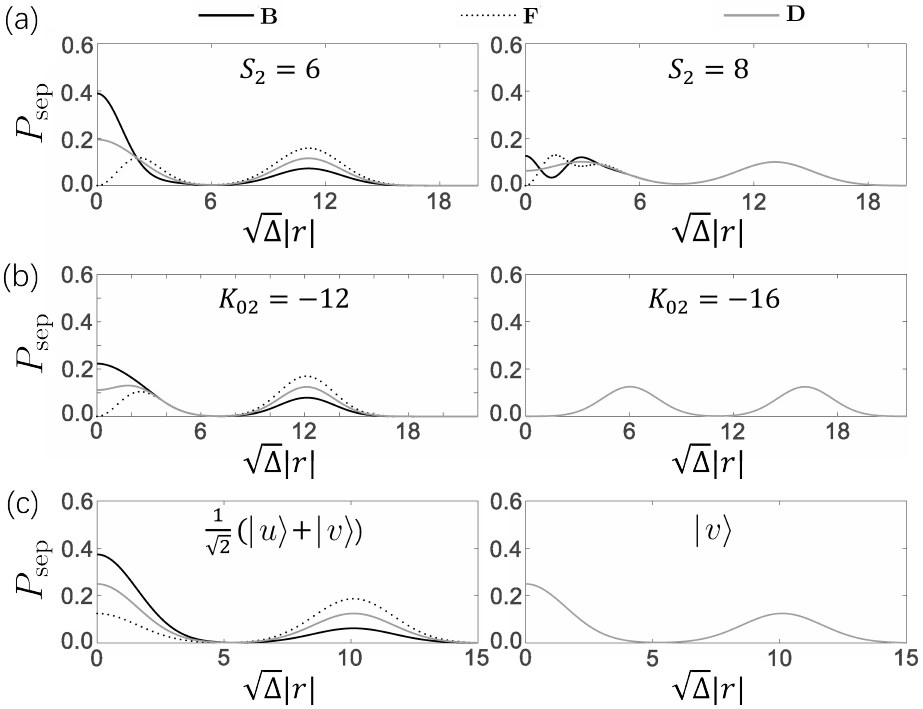}
	\caption{The scattered final states of bosons (B), fermions (F) and distinguishable particles (D) illustrate the impact of differences between the two incident particles on interference. The plots show the probability distribution for separation of particles when the dimensionless distance between the later outgoing wave packet $\Phi_{j}$ and the well equals $5$, with $j = 2$ for (a), $j = 1$ for (b), and $j = 1, 2$ for (c). (a) Initial distances from the well are different. (b) Incident energies are different. (c) Spin states are different. The parameters used to generate these plots are the same as in Fig.~\ref{Figure3}, except for those explicitly labeled.}
	\label{Figure6}
\end{figure}

In addition to the splitting ratio, differences in the incident energy, arrival time at the beam splitter, and spin state also affect interference. The resulting probability distribution for separation for fermions is no longer unimodal, as shown in Fig.~\ref{Figure6}. Therefore, all distribution curves --- for bosons, fermions, and distinguishable particles --- have two peaks. The areas of the two peaks correspond to the probabilities of the particles being detected on the same and opposite sides of the well, respectively, and the sum of these two probabilities equals $1$ \cite{sentence}.

In Fig.~\ref{Figure6}(a), the time delay changes with the initial distance $S_{2}$ from the potential well. As the time delay increases, the peaks for large separation coincide completely, and the peaks for small separation show interference fringes for identical particles, due to dispersion and overlap of wave packets. In Fig.~\ref{Figure6}(b), the difference in the incident energy varies with the value of $K_{02}$.  As the difference in energy increases, the distribution for all three types of particle develop an identical bimodal structure. (Both peaks move in the direction of increasing $|r|$ over time because the two particles have different speeds.) Different spin states in Fig.~\ref{Figure6}(c) cause the distribution of bosons and fermions to approach that of distinguishable particles until they coincide completely when the spin states are orthogonal. (The probability density for fermions at $|r| = 0$ can be non-zero only if their spin states are different.)

In general, the more different the initial states --- with respect to arrival time, energy, or spin --- the more all three probability distributions resemble that of distinguishable particles.  Interference effects disappear.

\section{CONCLUSION}

A famous phenomenon used in cutting-edge fields, the HOM effect is also suitable for undergraduate teaching. Compared with second quantization \cite{zongshu}, the approach described in this paper is more concrete and easier to visualize, and within the reach of undergraduates. The model shows the time evolution of all the important features in two-particle interference. Through comparison, it can help students to better understand the exchange symmetry of identical particles. Therefore, it can be a useful supplement to standard quantum mechanics textbooks. The interference process in Fig.~\ref{Figure3} can be animated to show the HOM effect \cite{SM}, or the story line can be designed into a seminar about identical particles. Teachers can design some problems based on our model and provide codes to students to explore different scattering scenarios. To promote discussion and cooperation, students are divided into small groups to work together and solve part of the problems. For example, some groups could explore bosons, some fermions, and some distinguishable particles. They could compare and contrast the results of their simulations, and report their findings and experiences to the class through a seminar.

\section{Supplementary Material}

Please click on this link to access the supplementary material, which includes the Octave codes for single particle scattering and two particle interference. The codes for generating Fig.~\ref{Figure1} and animating Fig.~\ref{Figure3} are provided. An animation of Fig.~\ref{Figure3} as a GIF file is also available. Print readers can see the supplementary material at [DOI to be inserted by AIPP]

\begin{acknowledgments}

 Z. J. Deng is grateful to Florian Marquardt and Ming-Qiu Huang for useful discussions. This work is supported by the National Natural Science Foundation of China under Grants No. 11574398, 12174448, and the Teaching Reform Research Project of Regular Higher Education Institutions in Hunan Province under Grant No. 202401000281. We also thank the developers of the free software GNU Octave \cite{software}, which was used for plotting and animation.

\end{acknowledgments}

\section{Author Declarations}

The authors have no conflicts to disclose.


\begin{thebibliography}{23}
\bibitem{sentence} The areas of the left peaks in Fig.~\ref{Figure6} are in good agreement with Eq.~(\ref{eq23}). For example, for the subplot with $S_{2} = 6$, the area of the left peaks obtained by numerical integrations for bosons, fermions, distinguable particles are $0.683$, $0.313$, $0.498$, respectively, while the corresponding values evaluated by Eq.~(\ref{eq23}) are $0.684$, $0.316$, $0.500$, respectively.
\end{thebibliography}
\end{document}